\documentclass[aps,floatfix,pra,reprint,superscriptaddress]{revtex4-1}
\usepackage{amsmath,amssymb,graphicx,mathtools}
\usepackage{float}

\usepackage[T1]{fontenc}

\usepackage{color}
\usepackage{colortbl}
\usepackage{nicefrac}
\usepackage[hypertexnames=false,colorlinks,urlcolor=blue,citecolor=blue]{hyperref}
\usepackage{natbib}
\usepackage[capitalize]{cleveref}

\makeatletter
\DeclareRobustCommand{\cev}[1]{%
  \mathpalette\do@cev{#1}%
}

\makeatother

\newcommand{\eg}{\emph{e.g.,~}}
\newcommand{\ie}{\emph{i.e.,~}}
\newcommand{\etal}{\emph{et al.}}

\newcommand{\nn}{\nonumber}

\newcommand{\dg}{^\dagger}
\newcommand{\bra}[1]{\langle #1|}
\newcommand{\ket}[1]{|#1\rangle}
\newcommand{\tr}{{\rm tr}}
\newcommand{\expect}[1]{\langle #1\rangle}

\makeatletter
\newcommand{\vast}{\bBigg@{4}}
\newcommand{\Vast}{\bBigg@{5}}
\makeatother

\newcommand{\snlca}{Quantum Algorithms and Applications Collaboratory, Sandia National Laboratories, Livermore, CA 94550, USA}
\newcommand{\cornell}{School of Applied and Engineering Physics, Cornell University, Ithaca, NY 14853, USA}
\newcommand{\unm}{Center for Quantum Information and Control (CQuIC), Department of Physics and Astronomy, University of New Mexico, Albuquerque, NM 87131, USA}
\newcommand{\lanl}{Theoretical Division, Los Alamos National Laboratory, Los Alamos, NM 87545, USA}

\newcommand{\papertitle}{Quantum computer-enabled receivers for optical communication}

\begin{document}
\title{\papertitle}
\author{John Crossman}
\affiliation{\snlca}
\affiliation{\cornell}
\author{Spencer Dimitroff}
\affiliation{\snlca}
\affiliation{\unm}
\author{Lukasz Cincio}
\affiliation{\lanl}
\author{Mohan Sarovar}
\email{mnsarov@sandia.gov}
\affiliation{\snlca}
\date{\today }

\begin{abstract}
Optical communication is the standard for high-bandwidth information transfer in today's digital age. The increasing demand for bandwidth has led to the maturation of coherent transceivers that use phase- and amplitude-modulated optical signals to encode more bits of information per transmitted pulse. Such encoding schemes achieve higher information density, but also require more complicated receivers to discriminate the signaling states. In fact, achieving the ultimate limit of optical communication capacity, especially in the low light regime, requires coherent joint detection of multiple pulses. Despite their superiority, such joint detection receivers are not in widespread use because of the difficulty of constructing them in the optical domain. In this work we describe how optomechanical transduction of phase information from coherent optical pulses to superconducting qubit states followed by the execution of trained short-depth variational quantum circuits can perform joint detection of communication codewords with error probabilities that surpass all classical, individual pulse detection receivers. Importantly, we utilize a model of optomechanical transduction that captures non-idealities such as thermal noise and loss in order to understand the transduction performance necessary to achieve a quantum advantage with such a scheme. We also execute the trained variational circuits on an IBM-Q device with the modeled transduced states as input to demonstrate that a quantum advantage is possible even with current levels of quantum computing hardware noise.
\end{abstract}
\maketitle

Quantum transduction is the task of converting quantum information from one carrier to another, with these carriers usually being degrees of freedom (DOF) at different energy scales; \eg superconducting qubits and optical photons. Traditionally, quantum transduction has been viewed as an element of quantum networking, enabling the connection of distributed quantum computers. However, another perspective is that quantum transduction, when the destination is one or more qubits in a scalable quantum computing platform, is a way to get unknown quantum states into a quantum computer. These states can then be used as input for a quantum computation, and when the transduction source is the electromagnetic (EM) field, this enables universal coherent processing of quantum information encoded in an EM field. With this perspective, quantum transduction coupled with quantum computing creates radically new modalities for sensing and detection of information in an EM field, a concept that we call \emph{quantum computational imaging and sensing (QCIS)} \cite{sarovar_qcis}. This term reflects the fact that the computing element and the sensing element are inexorably linked and cannot be separated, and extends the concept of computational imaging and sensing from the classical domain \cite{Spiegel_1996, bhandari_2022}. We note that a similar concept has been discussed in Ref. \cite{conlon_approaching_2023}.

In this work, we define and analyze a particular application of the QCIS concept: quantum receivers for higher rate coherent optical communication. This application is particularly useful for illustrating the advantage and utility of QCIS because it allows for quantitative comparison of performance against well-understood limits of conventional (classical) receivers. We perform this comparison and identify regimes of quantum advantage (as defined in \cref{sec:demo}) in the presence of non-idealities in the quantum transduction and quantum computation stages.

\begin{figure}[t]
\includegraphics[width=1\linewidth]{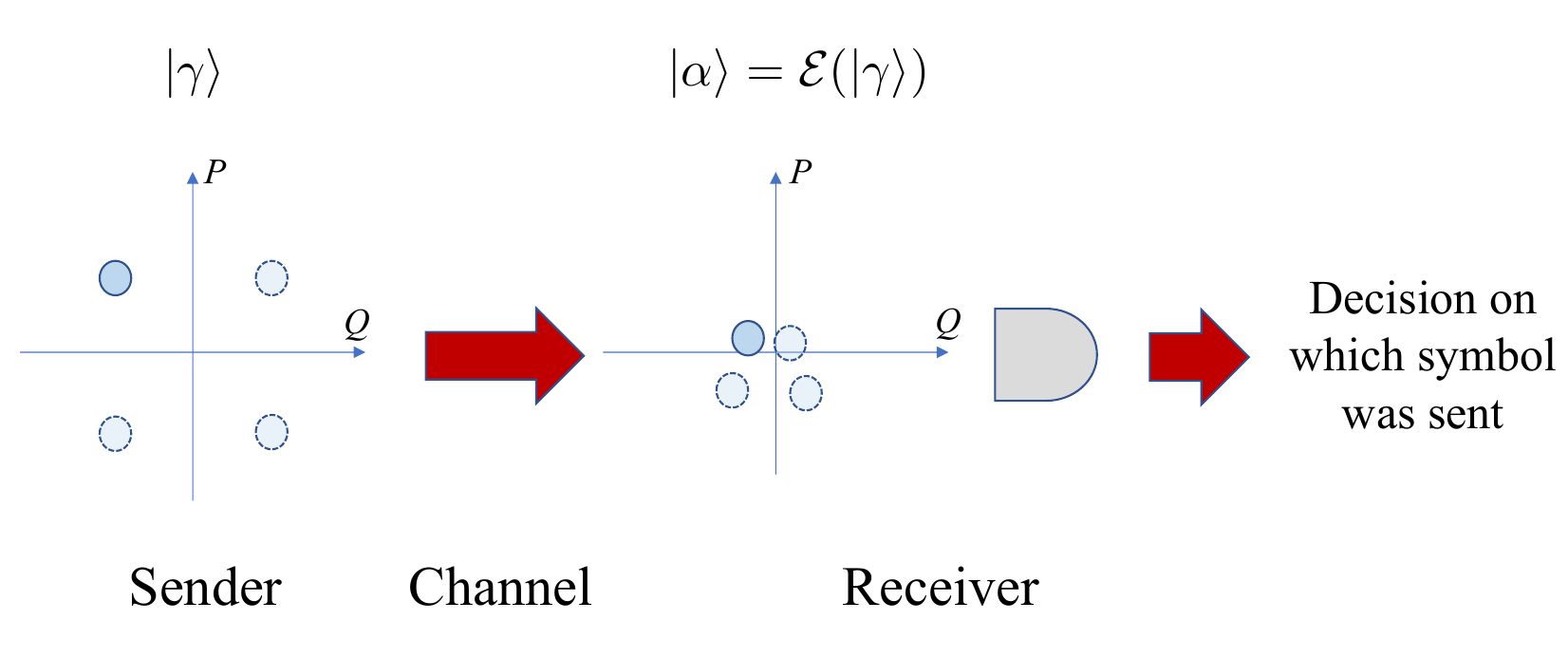}
	\caption{Schematic of coherent communication. The sender prepares one of $K$ coherent states (forming a constellation) corresponding to $K$ possible symbols to communicate -- in this example, $K=4$, which is known as quadrature phase shift keying (QPSK). In each use of the channel, the prepared state is sent to the receiver. As a result of channel loss and other non-idealities the states arriving at the receiver are distorted. The receiver measures the received pulses and attempts to make a decision about which coherent state was sent.\label{fig:coh_comm}}
\end{figure}

First, we introduce the application domain. Optical communication forms the backbone of the modern information age. Despite the staggering increase in communication bandwidth enabled by optical communication \cite{Agarwal_2010}, the ever-growing demand for bandwidth has led to the deployment of coherent optical communication networks. Coherent communication utilizes phase and amplitude degrees of freedom of an optical pulse, as opposed to just intensity, to squeeze more (classical) information into each pulse. Through phase and amplitude modulation of laser pulses the transmitter encodes information using one of the coherent states forming the communication \emph{constellation}, see \cref{fig:coh_comm}. The task of the receiver is to identify the transmitted state from the possible states in the constellation. Due to the non-orthogonality of coherent states there is always a finite probability of error associated with this task, even in the absence of non-idealities such as transmission loss and noise.

The ultimate limit to classical communication capacity when using a quantum state constellation is given by the Holevo bound on mutual information between sender and receiver, which is only a function of the quantum states received by the receiver \cite{holevo_1998, Schumacher_Westmoreland_1997}:
\begin{align}
    \mathcal{I} \leq \chi \equiv S(\rho) - \sum_i p_i S(\rho_i),
    \label{eq:chi}
\end{align}
with $\mathcal{I}$ being the mutual information, \(\rho= \sum_i p_i\rho_i\) and \(S(\cdot)\) the von Neumann entropy function. Here, \(\{\rho_i, p_i\}\) are the received quantum states and their prior probabilities. A classical receiver that measures the received pulses one at a time and decodes the resulting classical bits cannot attain the Holevo bound, regardless of the channel code utilized and generally, even if the measurements are performed adaptively. Instead, joint detection receivers (JDRs) that collectively measure multiple pulses and decode the resulting \emph{codeword} are required to approach the Holevo bound \cite{holevo_1998,Schumacher_Westmoreland_1997,Guha_2011}. This fact is sometimes known as the superadditivity of classical-quantum channel capacity \footnote{''Classical-quantum'' denotes that the aim is to send classical information over a quantum channel.} since the capacity achieved with collective measurements on a \emph{codeword} consisting of several pulses exceeds the sum of the capacities achievable by measuring each pulse separately. Despite the superadditivity of classical-quantum capacity, JDRs are not in widespread use since designing and constructing optimal JDRs is challenging. Theoretical progress has been made in this area recently \cite{Guha_IEEE_2011,Guha_2011,Wilde_2012,Guha_2012,Takeoka_2013}, but optical implementations of JDRs remain challenging, with the notable exception of the proof-of-principle implementation in Ref. \cite{Chen_2012}.

In this work, we address this application domain within the framework of QCIS. We utilize a model of optomechanical quantum transduction to transfer information from optical pulses to qubits and design a quantum computation on these qubits to jointly discriminate a transmitted codeword. The transduction model includes sources of noise and loss, and our analysis reveals regimes where this setup can surpass optimal single-pulse receivers. We note that recent work by Delaney {\etal} \cite{Delaney_2021} considered a very similar problem to that studied in this work, with two notable differences: (i) in the following we will develop a more complete transduction model than that considered by Delaney \etal, including the inclusion of realistic noise sources that impact performance, and (ii) the quantum computation step they employed was based on a message-passing decoding algorithm \cite{Rengaswamy_2020}, while we demonstrate that codeword states can be discriminated by variational quantum circuits, which could be more suitable for execution on noisy intermediate-scale quantum (NISQ) devices and when the transduction is non-ideal. 

The remainder of the article is organized as follows. In \cref{sec:trans} we detail the model of transduction that we use and how it enables transfer of optical coherent state information to superconducting qubits. In \cref{sec:variational} we outline the variational quantum circuit approach for performing quantum computations to discriminate codeword states. Then in \cref{sec:demo} we demonstrate the concept with numerical simulations and quantify regimes of quantum advantage. In \cref{sec:expt} we demonstrate a small-scale receiver on a cloud-based IBM-Q device by mimicking the input states that would result from the optomechanical transduction model. Finally, in \cref{sec:conc} we conclude with a discussion of the results and future work.

\section{Transduction of coherent states}
\label{sec:trans}
We base our analysis on arguably the most mature deterministic quantum transduction platform: optical to microwave frequency transduction through optomechanical systems. Several variations of this transduction mechanism have been demonstrated \cite{Andrews_2014,Higginbotham_2018,Forsch_2020,Mirhosseini_2020,Hease_2020,Witmer_2020,Brubaker_2022,Sahu_2022}. We begin with the theoretical model for this transduction platform presented by Tian and Wang \cite{Tian_Wang_2010}.
The goal will be to transfer an arriving optical coherent state (assumed to be in a well-defined spatial mode) into the optical cavity, and then transfer information about that coherent state (particularly the phase, which often encodes the classical information being transmitted) into the microwave cavity mode. Then the interaction between this microwave mode and a superconducting qubit will be engineered to transfer phase information into the qubit state.

The Hamiltonian describing this model is:
\begin{align}
	H &= \hbar\omega_1 b_1\dg b_1 + \hbar\omega_2 b_2\dg b_2 + \hbar\omega_3 b_3\dg b_3 + \frac{\hbar\Omega_q}{2} \sigma_z \nn \\
	&~~- \hbar g_1 b_1\dg b_1 q_2 - \hbar g_3 b_3\dg b_3 q_2 + \hbar\chi(b_3 \sigma^+ + b_3\dg \sigma^-) ,
	\label{eq:trans_H}
\end{align}
where \(b_i\) for \(i=1,2,3\) are annihilation operators for the principal mode of the optical cavity, mechanical oscillator and microwave cavity, respectively, and \(q_2 = \nicefrac{1}{\sqrt{2}}(b_2+b_2\dg)\). The first line describes the free Hamiltonians for all DOF, with \(\omega_1 > \omega_3 > \omega_2\) being the relevant energy scales. The second line describes the coupling between elements. The first two terms describe coupling of both of the cavities to the mechanical oscillator through the standard optomechanical coupling (the occupation of either cavity displaces the mechanical mode through radiation pressure forces). The third term describes the coupling between the microwave mode and qubit, described through a Jaynes-Cummings interaction since we assume the mode and qubit are close to resonance (\(\omega_3 \approx \Omega_q\)). In the following, we assume that all the couplings, \(g_1,g_3,\chi\), are tunable.

In addition to this Hamiltonian, to model the system dynamics we need to include dissipative terms. As a result, the evolution of the density matrix for the combined four DOF system, \(\varrho(t)\), follows the master equation
\begin{align}
	\dot{\varrho}(t) = -\nicefrac{i}{\hbar}[H,\varrho(t)] + \mathcal{L}_1\varrho(t) + \mathcal{L}_3\varrho(t) + \mathcal{D}\varrho(t),
	\label{eq:dissipative}
\end{align}
with
\begin{align}
	\mathcal{L}_j\varrho &= \kappa_j(b_j \varrho b_j\dg - \nicefrac{1}{2}b_j\dg b_j \varrho - \nicefrac{1}{2}\varrho b_j\dg b_j), \nn \\
	 \mathcal{D}\varrho &= \gamma(\bar{n}+1)(b_2\varrho b_2\dg - \nicefrac{1}{2}b_2\dg b_2 \varrho - \nicefrac{1}{2}\varrho b_2\dg b_2) \nn \\
	 & ~~~~~+ \gamma\bar{n}(b_2\dg\varrho b_2 - \nicefrac{1}{2}b_2 b_2\dg \varrho - \nicefrac{1}{2}\varrho b_2 b_2\dg).
\end{align}
\(\mathcal{L}_j\) model loss from each of the cavities at rates \(\kappa_j\), and \(\mathcal{D}\) models thermal damping and excitation of the mechanical oscillator. \(\gamma\) is the coupling rate to the thermal reservoir and \(\bar{n} = \nicefrac{1}{(e^{\hbar\omega_2/k_B T}-1)}\) is the average occupation of the temperature \(T\) reservoir at the mechanical oscillation frequency. We assume that the qubit decoherence is negligible during the transduction process.

The parameters entering this model vary according to the physical realization of the optomechanical system. In this work we base our analysis on the devices reported in Refs. \cite{Andrews_2014,Higginbotham_2018}, and list the fiducial values for the parameters used in simulations and analysis in \cref{app:parameters}.

\subsection{Linearization through parameteric driving}
The first step in using the dynamics of the system to transduce coherent states from the optical domain to the microwave frequency qubit is to linearize the optomechanical interactions \cite{Tian_Wang_2010}. In order to do this, we add coherent drives to both the optical and microwave cavities at frequencies \(\nu_1, \nu_3\):
\begin{align}
	H_d(t) = \hbar\sum_{j=1,3} E_je^{-i\nu_j t}b_j\dg + E^*_j e^{i\nu_j t} b_j.
\end{align}
The drive frequencies are red-detuned from both cavities by approximately the frequency of the mechanical oscillator; \ie \(\nu_j = \omega_j - \Delta_j\), with \(\Delta_j \approx \omega_2\). The Hamiltonian for the system in a rotating frame with respect to these drives is:
\begin{align}
	\tilde{H} &= \hbar\Delta_1 b_1\dg b_1 + \hbar\omega_2 b_2\dg b_2 + \hbar\Delta_3 b_3\dg b_3 \nn \\
	&~~~- \hbar g_1 b_1\dg b_1 q_2 - \hbar g_3 b_3\dg b_3 q_2 + \hbar\sum_{j=1,3} E_jb_j\dg + E_j^* b_j \nn \\
	&~~~+ H_q,
	\label{eq:trans_H_rot}
\end{align}
where \(H_q \equiv \frac{\hbar\Omega_q}{2}\sigma_z + \hbar\chi(b_3\sigma^+ + b_3\dg \sigma^-)\). Assuming the qubit coupling \(\chi\) is turned off, the Heisenberg equations of motion for the mode operators, including the dissipative dynamics in \cref{eq:dissipative}, are:
\begin{align}
	\dot{b}_j &= -i(\Delta_j-g_jq_2)b_j -i E_j -\frac{\kappa_j}{2}b_j, \quad j=1,3 \nn \\
	\dot{b}_2 &= -i\omega_2 b_2 +\frac{ig_1}{\sqrt{2}}b_1\dg b_1 +\frac{ig_3}{\sqrt{2}}b_3\dg b_3 -\frac{\gamma}{2}b_2.
	\label{eq:eom}
\end{align}
The driving and dissipation result in steady-states of these modes, which are given by:
\begin{align}
	\mathcal{B}_j &= \frac{-iE_j}{\nicefrac{\kappa_j}{2} + i(\Delta_j-g_j\mathcal{Q}_2)}, \quad j=1,3 \nn \\
	\mathcal{B}_2 &= \frac{\nicefrac{ig_1}{\sqrt{2}}|\mathcal{B}_1|^2 + \nicefrac{ig_3}{\sqrt{2}}|\mathcal{B}_3|^2}{i\omega_2 + \nicefrac{\gamma}{2}},
	\label{eq:steady_states}
\end{align}
where \(\mathcal{Q}_2=\nicefrac{1}{\sqrt{2}}(\mathcal{B}_2+\mathcal{B}_2^*)\) is the steady-state of the operator \(q_2\). We expand the mode operators around these steady states, \eg \(b_j = \mathcal{B}_j + \delta b_j\), defining new annihilation operators \(\delta b_j\) that quantize fluctuations around the steady-states for each DOF. Writing the equations of motion, \cref{eq:eom}, in terms of these expansions and neglecting terms quadratic or higher in \(\delta b_j\) yields a linearized approximation to the evolution:
\begin{align}
	\dot{\delta b}_j &= -i(\Delta_j -g_j\mathcal{Q}_2)\delta b_j + i g_j \mathcal{B}_j\delta q_2 - \frac{\kappa_j}{2} \delta b_j, \quad j=1,3\nn \\
	\dot{\delta b}_2 &= -i\omega_2\delta b_2 + \sum_{j=1,3} \frac{ig_j}{\sqrt{2}}(\mathcal{B}^*_j \delta b_j + \mathcal{B}_j \delta b_j\dg) - \frac{\gamma}{2}\delta b_2, \nn
\end{align}
which is generated by the linearized Hamiltonian (in the rotating frame),
\begin{align}
	\tilde{H}_L &= \hbar\sum_{j=1,3} (\Delta_j - g_j\mathcal{Q}_2) \delta b_j\dg\delta b_j - g_j(\mathcal{B}_j^* \delta b_j + \mathcal{B}_j \delta b_j\dg)\delta q_2 \nn \\
	&~~+ \hbar\omega_2 \delta b_2\dg \delta b_2,
	\label{eq:lin_ham}
\end{align}
and the same dissipative generators as in \cref{eq:dissipative} (redefined by replacing each \(b_j\) with \(\delta b_j\)).

To obtain the final form of the linearized interaction Hamiltonian we will assume that \(|g_j\mathcal{B}_j|\ll \omega_2\), and hence drop the counter-rotating terms in the interactions above, \ie \((\mathcal{B}_j^*\delta b_j + \mathcal{B}_j\delta b_j\dg) \delta q_2 \rightarrow (\mathcal{B}_j^*\delta b_j \delta b_2\dg + \mathcal{B}_j\delta b_j\dg \delta b_2)\), to get a beam-splitter interaction between the shifted modes.

As a result of the parameteric drive, we have a beam-splitter interaction between all three (shifted) oscillator degrees of freedom, which is ideal for state transfer. In addition, note that (i) the shifted modes acquire a frequency shift \(\Delta_j - g_j \mathcal{Q}_2\) that is \emph{a priori} calculable given knowledge of the system parameters, and (ii) the coupling between modes has been modified from the pure optomechanical coupling rate \(g_j\) to \(G_j \equiv g_j|\mathcal{B}_j|\). This is typically an enhancement since in good operating regimes, \(|\mathcal{B}_j|>1\). Finally, since the linearized Hamiltonian is in terms of the shifted operators \(\delta b_j\), it is important to note that the states we will consider are states that sit atop the displaced vacuum for all three DOFs: \(\ket{\mathcal{B}_j}\) for \(j=1,2,3\).

\subsection{Optical-to-microwave transduction protocol}
Given the linearized model derived above, there are several possible protocols for transferring optical coherent states to the microwave cavity, including sequential swapping of states from a cavity to the resonator and \emph{vice versa} \cite{Tian_Wang_2010}, adiabatic transfer \cite{Tian_2012}, and a hybrid version of these \cite{Wang_Clerk_2012}. In the following, we will base our analysis on the sequential swap protocol, which sequentially transfers the incoming coherent state pulse into the optical cavity, optical cavity to mechanical oscillator, and then mechanical oscillator to microwave cavity by tuning \(g_j\). This protocol is essentially the same as the one developed by Tian and Wang \cite{Tian_Wang_2010} and is summarized in \cref{app:protocol}.

The first step in this protocol inputs the arriving coherent state pulse \(\ket{\alpha}\) into the optical cavity. Here, \(\alpha \in \mathbb{C}\) incorporates any channel loss between the sender and receiver. The carrier frequency of the incoming pulse is resonant with the optical cavity, and the Heisenberg equation of motion describing the dynamics of the cavity mode in a frame rotating at \(\omega_1\) is:
\begin{align}
	\dot{b}_1(t) = -\frac{\kappa_1}{2}b_1(t) + \sqrt{\kappa_1}b_{\rm in}(t),
\end{align}
where \(b_{\rm in}(t)\) is the input field. This linear interaction between the cavity mode and the input field means that an input coherent state is transferred into a coherent state in the cavity, and solving for the expectation of the cavity mode yields:
\begin{align}
	\expect{b_1(t)} = \sqrt{\kappa_1}e^{-\nicefrac{\kappa_1 t}{2}}\int_0^t {\rm d}\tau e^{\nicefrac{\kappa_1 \tau}{2}}\expect{b_{\rm in}(t)},
\end{align}
where we have assumed the cavity is unoccupied (vacuum) at \(t=0\). We consider a coherent state input, \(\expect{b_{\rm in}(t)} = \alpha f(t)\), where \(f(t)\) is the pulse profile. For a square pulse of temporal width \(T\) and a constant input-output coupling, \(\kappa_1\), the cavity mode at time \(t=T\) is \(\expect{b_1(T)} = \nicefrac{\alpha}{\sqrt{\kappa_1}}\left(2(1-e^{-\kappa_1T/2})\right)\). Thus the coherent state is transferred into the optical cavity after suffering some \emph{insertion loss}. The factor \(\eta_{\rm in} = \left(\nicefrac{2(1-e^{-\kappa_1T/2})}{\sqrt{\kappa_1}}\right) \) ($< 1$ in practice) quantifies the insertion, or input, efficiency. This efficiency can be tuned, and made close to unity, by either engineering the pulse profile \cite{Wang_2011} or by dynamically tuning the input-output coupling \(\kappa_1\) \cite{Chatterjee_2022}. In addition, any traditional pulse-by-pulse receiver will also incur some insertion loss (that is similarly tunable by a variety of techniques). As a result of this tunability, and in order to compare the unique features of transduction-based receivers to traditional optical receivers, we ignore this insertion loss, and simply assume that a coherent state \(\ket{\beta = \alpha}\) is initialized in the optical cavity at a known time. Our further analysis will look at the effect of \(|\beta|^2\) or received mean photon number (RMPN) on JDR performance. When comparing to traditional optical receivers we assume they suffer no insertion loss as well, and have access to the same RMPN.

Since the initial state in the optical cavity is Gaussian and the transfer dynamics are linear (governed by a quadratic Hamiltonian and linear dissipative operators), we know the transferred state in the microwave cavity is also Gaussian. In fact, Wang and Clerk have derived the analytic form of the transferred Gaussian state under the linearized model \cite[Appendix B]{Wang_Clerk_2012}. Given an initial coherent state \(\ket{\beta}\) in the optical cavity, a thermal state with thermal occupation \(\bar{n}_0\) of the mechanical oscillator (\ie we do not assume perfect ground state cooling of the oscillator) and the vacuum state in the microwave cavity, the state transferred to the microwave cavity by the protocol is a displaced thermal state $\varrho( \bar{n}_{\rm tr}, \mathcal{B}_3+\sqrt{\eta_{\rm tr}}\beta)$, with
\begin{align}
	\varrho( \bar{n}, \alpha_0) \equiv \frac{1}{\pi \bar{n}}\int {\rm d}^2\alpha e^{-\frac{|\alpha - \alpha_0|^2}{\bar{n}}} \ket{\alpha}\bra{\alpha}.
	\label{eq:disp_thermal}
\end{align}
This is a Gaussian state with (quadrature) mean and covariance matrix \((\sqrt{2}\textrm{Re}(\mathcal{B}_3+\sqrt{\eta_{\rm tr}}\beta), \sqrt{2}\textrm{Im}(\mathcal{B}_3+\sqrt{\eta_{\rm tr}}\beta))^{\sf T}\) and \(V = (\bar{n}_{\rm tr}+\nicefrac{1}{2})I_2\), respectively. Therefore, the transduction from the optical cavity to the microwave cavity results in attenuation and heating of the input state. The attenuation and heating parameters, \(\eta_{\rm tr}\) and \(\bar{n}_{\rm tr}\), are expressed and plotted in terms of the physical model parameters in \cref{app:protocol}, and
\cref{tab:params} shows these parameters for the fiducial device parameters considered in \cref{app:parameters} at two operating temperatures.

\begin{table}[t]
\begin{tabular}{l|c|c}
 & \(\bar{n}_{\rm tr}\) & \(\eta_{\rm tr}\) \\
 \hline
 \(T=1\)K & 1.8 & 0.924 \\
 \(T=1\)mK & 0.001 & 0.924
\end{tabular}
\caption{Transduction heating and loss parameters for fiducial transduction model parameters in \cref{app:parameters} at two operating temperatures. \label{tab:params}}
\end{table}

While the device parameters we use to arrive at the loss and heating rates in \cref{tab:params} are optimistic, they are not out of reach of modern experimental platforms. 
Experimental demonstrations of optomechanical transducers have operated at temperatures ranging from $4$K \cite{Andrews_2014} to as low as $7$mK \cite{Hease_2020, Witmer_2020}, with the latter achieved through operation inside optical-access dilution refrigerators. In a continuously operated electro-optomechanical transducer, bidirectional conversion efficiencies as high as 47\% have been achieved with 3.2 input-referred added noise photons in the upconversion process (microwave-to-optical) \cite{Brubaker_2022}. Alternatively, input-referred added noise levels as low as 0.16 photons in upconversion have been obtained in a pulsed nonlinear electro-optic platform with 8.7\% bidirectional efficiency \cite{Sahu_2022}. The gap between these efficiencies and heating rates can be attributed to two factors. Firstly, our model ignores technical sources of imperfection, such as laser noise, optical pump absorption heating, and intrinsic materials loss, which can be minimized in principle. Secondly, as explained in \cref{app:parameters}, to be within the strongly coupled regime that enables high-fidelity optical-to-microwave transduction \cite{Wang_Clerk_2012}, we assume much smaller optical and microwave cavity linewidths ($\kappa_1, \kappa_3$) than demonstrated thus far in optomechanical devices. However, we note that such narrow linewidth cavities are well within reach of modern devices \cite{eerkens2015optical,planz2023membrane}.

The state in the microwave cavity after the cavity transduction steps is \cref{eq:disp_thermal}. However, note that the signal we care about is only a small part of the displacement \(\mathcal{B}_3+\sqrt{\eta_{\rm tr}}\beta\), since typically \(|\mathcal{B}_3|\gg |\beta|\). We can compensate for this steady state field offset, which recall is necessary to linearize the optomechanical interactions, by applying a compensation drive to the microwave cavity after the transduction step. Specifically, the expectation of the microwave cavity annihilation operator at time \(t\) after a drive \(a_{\rm in}\) is applied is \cite{Walls_Milburn_2007}
\begin{align}
	\expect{b_3(t)} = e^{-\frac{\kappa_3 t}{2}}\expect{b_3(0)} + \sqrt{\kappa_3}e^{-\frac{\kappa_3 t}{2}}\int_0^t d\tau e^{\frac{\kappa_3 \tau}{2}}\expect{a_{\rm in}(\tau)}. \nn
\end{align}
The initial state is \(\expect{b_3(0)} = \mathcal{B}_3+\sqrt{\eta_{\rm tr}}\beta\), and choosing the drive to be time-dependent and of the form \(\expect{a_{\rm in}(\tau)} = -e^{-\nicefrac{\kappa_3 \tau}{2}}\frac{\mathcal{B}_3}{\sqrt{\kappa_3}}\), yields at time \(t=1\), \(\expect{b_3(t=1)} = e^{-\frac{\kappa_3}{2}}\sqrt{\eta_{\rm tr}}\beta\). Therefore, the offset field can be removed at the cost of additional loss, \(\eta_{\rm tr} \rightarrow e^{-\kappa_3}\eta_{\rm tr}\). We assume \(\kappa_3\) is tunable and can be made arbitrarily small during this compensating drive (at the cost of increasing the drive power), and therefore ignore this additional loss in the following. 

\subsection{Microwave mode to qubit transduction}
The final step in the transduction chain is to transfer information about \(\beta\) to the qubit coupled to the microwave cavity mode. For this step we assume \(\chi\), the mode-qubit coupling, is turned on while \(G_1=G_3=0\). We also assume \(\chi\gg \kappa_3\), that the qubit is brought in resonance with the microwave mode, and that the coherent coupling between mode and qubit dominates any qubit decoherence as well. In this case we simply have evolution under the Jaynes-Cummings (JC) Hamiltonian \(\chi(b_3\sigma^+ + b_3\dg \sigma^-)\) (in an interaction picture with respect to \(\hbar\omega_3 b_3\dg b_3 + \frac{\hbar\Omega_q}{2}\sigma_z\)), of an initial state \(\rho_f(0) \otimes\rho_q(0)\). We set the qubit initial state to be the ground state, \(\rho_q(0)=\ket{g}\bra{g}\), and the JC interaction time can be optimized to maximize distinguishability between the transduced qubit states. This optimal time will have a complicated dependence on \(\bar{n}_{\rm tr}, \eta_{\rm tr}, \beta\) and \(\chi\), and in \cref{sec:JC} we explore this dependence and other aspects of the JC dynamics.

\begin{figure}[t]
\includegraphics[width=0.95\linewidth]{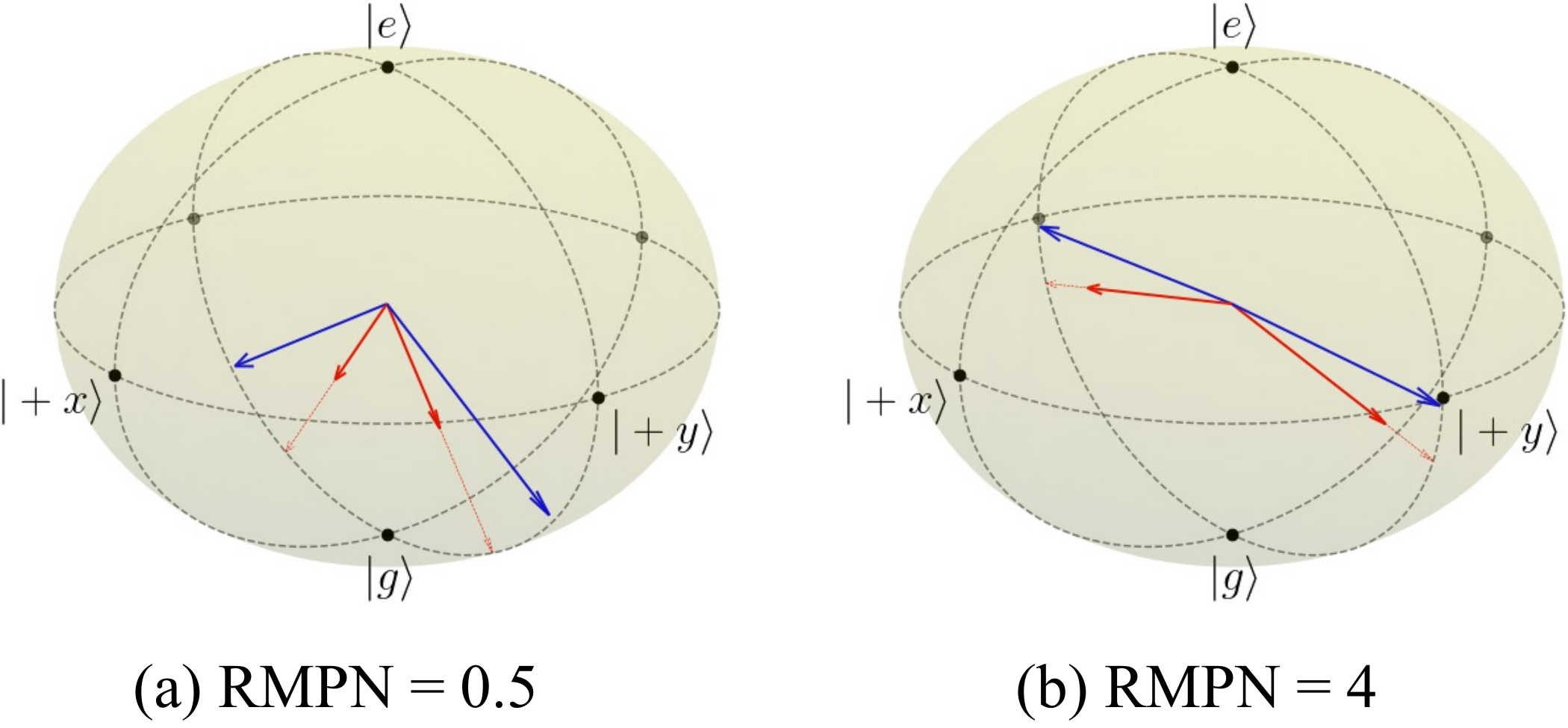}
	\caption{Qubit states resulting from transduction of BPSK states for two values of received mean photon number, with transduction under the fiducial model parameters in \cref{app:parameters}. The blue arrows correspond to low temperature (\(T=1\)mK, \(\bar{n}_{\rm tr} = 0.001\)) and the red arrows to high temperature (\(T=1\)K, \(\bar{n}_{\rm tr} = 1.8\)) transduction. The dashed red arrows extrapolate from the solid red arrows to the surface of the Bloch sphere and are a visual aid to show the reduction in Bloch vector length.\label{fig:qubits}}
\end{figure}

As an example, consider binary phase shift keying (BPSK) signaling, where the two possible initial states of the microwave mode are \(\rho_f(0)=\varrho(\bar{n}_{\rm tr},\pm \sqrt{\eta_{\rm tr}} \beta)\). In \cref{fig:qubits} we plot the two transduced qubit states on the Bloch sphere after numerical optimization of the JC interaction time for two values RMPN and the two operating temperatures considered (\cref{tab:params}). These figures illustrate that the field rotates the initial qubit state towards the \(X-Y\) plane on the Bloch sphere, with the amount of rotation dictated by the average number of photons in the transduced microwave mode, which is a function of the RMPN and transduction loss. Furthermore, transduction-induced heating reduces distinguishability of the qubits by making the states more mixed. Given a large enough RMPN, \eg \(|\beta|^2=4\), and low transduction-induced heating, the rotation induced by the two possible phases can result in nearly orthogonal states (\cref{fig:qubits}(b), blue arrows). The distinguishing feature of the BPSK states, \ie the phase arg(\(\beta\)), imprints itself onto the azimuthal phase of the qubit as $i \arg(\beta^*)$, yielding transduced qubit states with opposite phases. See \cref{sec:JC} for more details. 

\section{Variational circuits for state discrimination}
\label{sec:variational}
Once the optical coherent states used for communication have been transduced into qubits via the transduction mechanism detailed in \cref{sec:trans}, the task of the receiver becomes that of distinguishing between the possible qubit states. We define a codeword as a collection of received pulses that are each transduced into individual qubits. Ultimately achieving channel capacity requires joint decoding of asymptotically large codewords, but one can surpass classical receiver performance even with finite, small codewords. The quantum computation to distinguish the codewords can be constructed using existing decoding strategies as was done in Refs. \cite{daSilva_2013,Delaney_2021}. In contrast, in this work we utilize trained variational quantum circuits to discriminate the possible codewords. This strategy has a number of advantages: (i) it does not require a known decoding strategy, which is especially important when non-idealities of transduction are considered since good decoding strategies for ideal optical coherent state codewords may not port to decoding of imperfect, mixed-state qubit-encoded codewords, (ii) it is more compatible with NISQ computers since the variational ansatz can be chosen according to the circuit depths that can be executed with low error on a given device.

\begin{figure}[t!]
\includegraphics[width=1\linewidth]{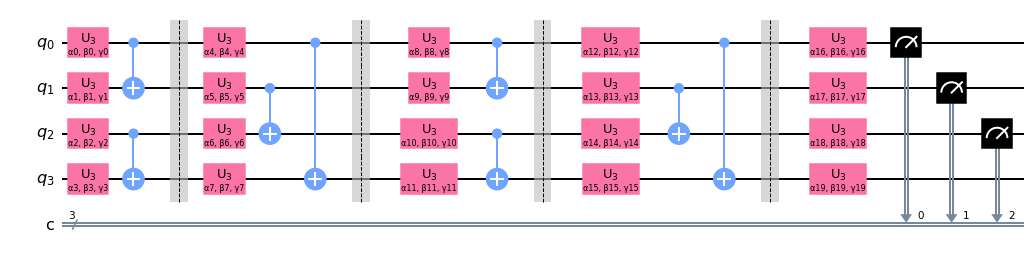}
	\caption{Four-qubit example of the variational circuit ansatz used in \cref{sec:demo}. Alternating layers of CNOT gates between neighboring qubits (assuming cyclic boundary conditions) conjugated by arbitrary single qubit gates are repeated. For four qubits, two time steps are required to execute CNOTs between all neighbors. Therefore, the example above contains two layers of the ansatz. At the end of the circuit \(\lceil\log_2(M)\rceil\) qubits (where \(M\) is the number of signaling states being distinguished) are measured and the bit string outcomes are assigned to each of the signaling states. The variational parameters are in the single qubit gates, each one having three angle parameters; \eg for the example shown here, there are 60 variational parameters. \label{fig:variational_circuit}}
\end{figure}

Variational quantum circuits consist of gates with tunable parameters that can be optimized for some task. Typically, the circuits take the form of some \emph{ansatz} that specifies the circuit structure. In this work we consider variational circuits with the structure illustrated in \cref{fig:variational_circuit}. The number of tunable layers (and hence variational parameters) can be varied, and we expect to require more layers as the number of codewords to be discriminated increases. Suppose \(M\) possible codewords are transmitted using \(n\) pulses that are transduced into \(n\) qubits (\(M \leq 2^n\)). We measure a fixed \(\lceil \log_2(M) \rceil\) of the qubits at the circuit output in the computational basis and assign one of the resulting bit strings to each codeword. Then the variational cost function that is maximized is
\begin{align}
	\mathcal{J}(\boldsymbol{\theta}) = \frac{1}{M}\sum_{i=1}^M p(b_i | \rho_i) =  \frac{1}{M}\sum_{i=1}^M \bra{b_i} U\dg(\boldsymbol{\theta})\rho_i U(\boldsymbol{\theta})\ket{b_i},
	\label{eq:perr}
\end{align}
where \(\boldsymbol{\theta}\) are the variational parameters of the circuit $U(\boldsymbol{\theta})$, and \(p(b_i|\rho_i)\) is the probability of measuring bit string \(b_i\) that corresponds to the codeword \(i\), when the input to the circuit is \(\rho_i\), the \(n\)-qubit state that encodes codeword \(i\). This cost function is just the average probability of successful decoding, assuming all codewords are sent with equal probability. We numerically train the variational circuits using a method derived from the qFactor optimizer~\cite{kukliansky2023qfactor}. The cost function is quadratic in gates, which makes the standard update described in Ref. ~\cite{kukliansky2023qfactor} numerically unstable. This is remedied by introducing a regularization factor $\beta$ as described in Sec.~IIIA of that paper. QFactor is applied to optimize over the variational ansatz in \cref{fig:variational_circuit} with varying numbers of layers (\ie circuit depths), but it can also optimize over unitaries as opposed to variational circuits. As a result, we also find system-size unitaries that maximize the cost function without assuming any circuit structure. Provided the optimization is successful, this gives us the largest possible cost function that can be achieved with a quantum circuit, and in \cref{sec:demo} we show the performance achieved by such optimized $n$-qubit unitaries as well as the optimized variational circuits. 

The chosen mapping between input states and output bit strings $\rho_i \rightarrow b_i$ can effect the average error probability if the number of variational parameters in the decoding circuit is small. However, given enough variational parameters, \ie enough layers in the variational ansatz, the cost function value becomes independent of this mapping since the circuit becomes expressive enough to implement the qubit permutations necessary to optimize $\mathcal{J}$. For small variational circuits this mapping can be optimized along with the circuit to further improve performance, although we do not do this here. 

Note that while the variational circuit training cost scales exponentially with codeword size, this is a one-off cost. For small codeword sizes \((n\lesssim 15-20)\) the training can be done numerically once the transduced qubit states are characterized. For larger codeword sizes we envision that the training is done with the quantum computing device itself evaluating the cost function.

\section{Demonstration of quantum computer-enabled joint detection receivers}
\label{sec:demo}
In this section we combine the coherent state transduction model and variational quantum processing model to demonstrate a joint detection receiver that exceeds the performance of all ``classical'' individual pulse receivers. We restrict ourselves to considering BPSK-based coherent communication, but extension to other communication constellations is straightforward.

For BPSK, the received optical states \(\ket{\pm \beta}\) result in two possible transduced single qubit states \(\rho_\pm\). We will study transduction with the fiducial parameters presented in \cref{app:parameters} and at temperatures \(T=1\)K and \(T=1\)mK, resulting in the transduction heating and loss parameters in \cref{tab:params}.

The achievable capacity for the optimal conventional receiver that decodes using pulse-by-pulse detection is $C_1 = 1 - h_2(\frac{1}{2}-\frac{1}{2}\sqrt{1-e^{-4|\beta|^2}})$ bits per pulse, where $h_2$ is the binary entropy function \cite{Guha_2011}. In contrast, as discussed in the Introduction, the asymptotically achievable capacity, enabled by joint detection of codewords is $C_\infty = \chi$ bits per pulse, see \cref{eq:chi}. Here, the Holevo quantity is computed using the state ensemble per pulse available at the receiver with equal prior probabilities for the symbols in the constellation. This capacity is achievable if one uses an error correction code with decoder based on joint measurement of codewords, \eg \cite{Rengaswamy_2020}, or using a codebook with $M=2^{nR}$ random $n$-bit codewords where each bit is encoded using one of the BPSK symbols. Here, $R$ is the rate of the code and if $R<C_\infty$ and the receiver attempts to optimally discriminate between the $M$ codewords, the probability of decoding error goes to zero as $n\rightarrow\infty$.

\begin{figure*}[t!]
 	\includegraphics[width=0.7\linewidth]{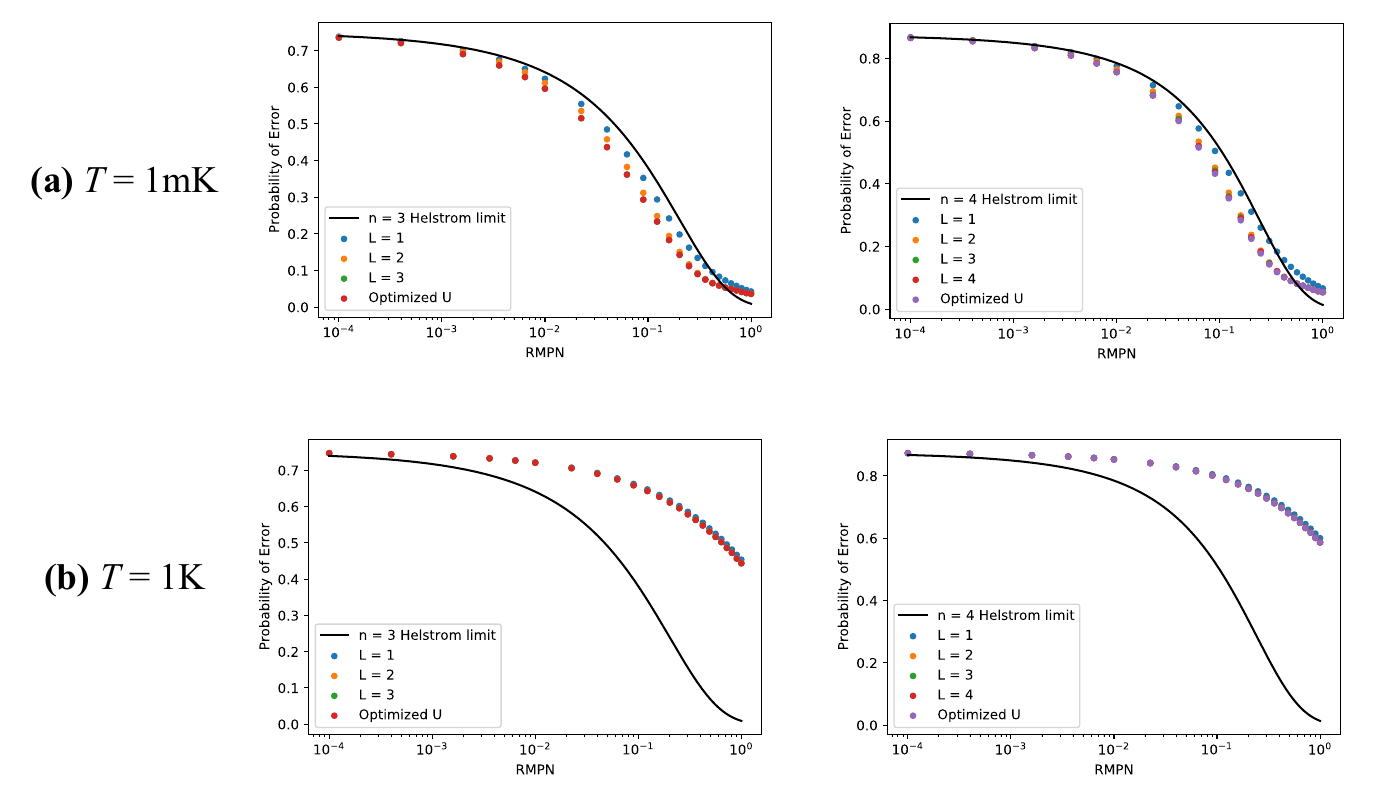}
	\caption{Average probability of error ($1-\mathcal{J}$) for decoding $M=2^{n-1}$ $n$-qubit codewords using optomechanical transduction and trained variational circuits for $n=3,4$. \textbf{(a)} shows results for transduction at low temperature ($1$mK), and \textbf{(b)} shows results for transduction at high temperature ($1$K). In all plots, the black curves are the relevant Helstrom limits that capture the best pulse-by-pulse receiver performance, as explained in the main text. The dots show the performance of the quantum computer-based JDR for varying circuit depths -- $L$ is the number of layers of the variational ansatz in \cref{fig:variational_circuit} that are optimized. The ``Optimized U'' dots correspond to $(1-\mathcal{J})$ achievable by optimized $n$-qubit unitaries.  \label{fig:n3n4}}
\end{figure*}

In the following, we will present the probability of decoding error for varying codeword lengths $n$, and in all cases, unless otherwise specified, we choose $M=2^{n-1}$. The recieved $n$-pulse codeword is transduced into $n$ qubits; \eg the length three optical codeword $\ket{+\beta}\otimes \ket{-\beta}\otimes \ket{-\beta}$ is transduced to $\rho_+\otimes \rho_-\otimes \rho_-$.

\begin{figure}[t!]
 	\includegraphics[width=1\linewidth]{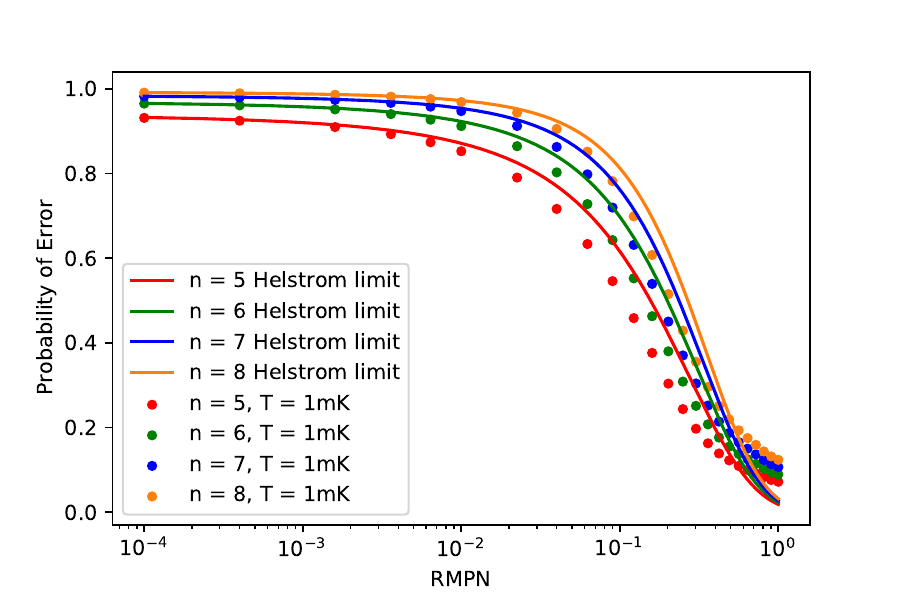}
	\caption{Average probability of error ($1-\mathcal{J}$) for decoding $M=2^{n-1}$ $n$-qubit codewords using optomechanical transduction and trained variational circuits for $n=5 - 8$. This figure only shows results for low temperature transduction, and the $(1-\mathcal{J})$ achieved by ideal, optimized $n$-qubit unitaries. The solid curves are the relevant Helstrom limits that capture the best pulse-by-pulse receiver performance, as explained in the main text. \label{fig:n5-8}}
\end{figure}

\cref{fig:n3n4} shows the average probability of decoding error, $(1-\mathcal{J})$ for codeword sizes $n=3,4$ as a function of the RMPN at the two transduction operating temperatures ($T=1$K and $T=1$mK). We focus on the region of low RMPN because this is where the greatest benefit from using a JDR is expected. We show the average probability of error achievable using variational circuit ans\"atze of varying depths, and an optimized $n$-qubit unitary. The black curve in all figures shows the \emph{n Helstrom limit}, which is the minimum error probability achievable when performing individual pulse-by-pulse detection in the optical domain with $n-1$ received BPSK pulses with the given RMPN (it is $n-1$ pulses as opposed to $n$ because with the JDR codewords we are communicating $n-1$ bits). Specifically, the $n$ Helstrom limit is
\begin{align}
p_{\rm n~Helstrom} = 1 - [1 - p_{\rm H}(\bar{n})]^{n-1},
\end{align}
where $\bar{n}$ is the RMPN and $p_{\rm H} = \frac{1}{2}-\frac{1}{2}\sqrt{1-e^{-4\bar{n}}}$ is the Helstrom bound on error probability per BPSK pulse. Achieving this ``classical bound'' requires using a Helstrom-bound saturating detector like the Dolinar receiver \cite{Dolinar_1973}.

\begin{figure}[t!]
 	\includegraphics[width=1\linewidth]{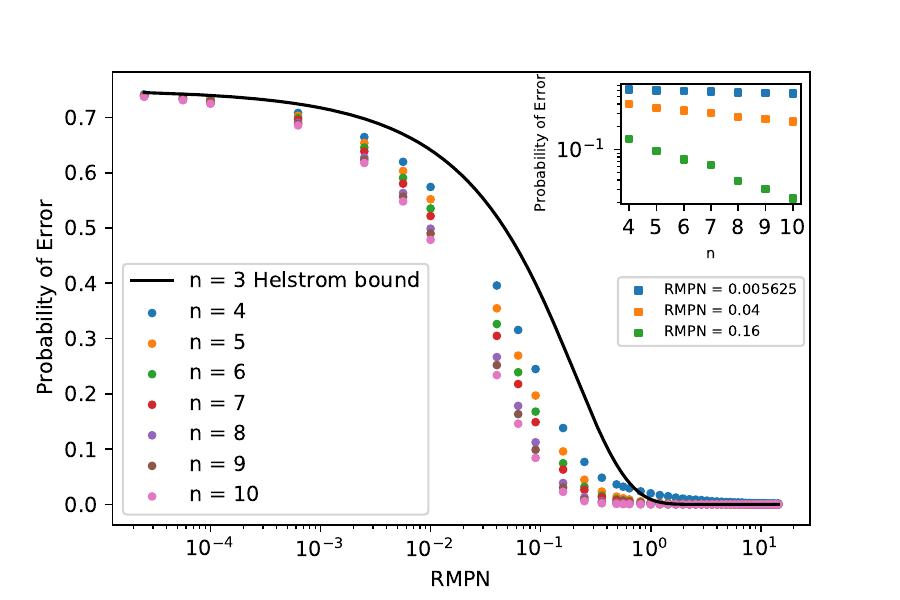}
	\caption{Average probability of error ($1-\mathcal{J}$) for decoding $M=4$ $n$-qubit codewords using optomechanical transduction and trained variational circuits for $n=4 - 10$. This figure only shows results for low temperature transduction, and the $(1-\mathcal{J})$ achieved by ideal, optimized $n$-qubit unitaries. The solid curve is the Helstrom limit that captures the best pulse-by-pulse receiver performance when decoding two pulses (transmitting 2 bits). The inset shows the trend of exponentially decreasing average probability of error with increasing codeword size, $n$, at various values of RMPN. \label{fig:codewords}}
\end{figure}

\cref{fig:n3n4}, although its only for small codeword sizes, reveals several interesting insights. First, it is clear from \cref{fig:n3n4}(b) that the amount of noise introduced by transduction at $T=1$K renders discriminating the codewords difficult, even in the ideal case with optimized $n$-qubit untaries. The JDR $(1-\mathcal{J})$ does not reduce below the relevant Helstrom limits for any values of RMPN. In contrast, for low temperature transduction, \cref{fig:n3n4}(a), there is a significant region of RMPN where the $(1-\mathcal{J})$ achieved by the JDRs improves over the Helstrom limit. Increasing the variational circuit depth allows reduction of $(1-\mathcal{J})$, however, for $n=3 (4)$, the average error probability almost saturates to values achievable with full unitary optimization by $L=3 (4)$ already.

In \cref{fig:n5-8} we show how the average probability of error behaves (as a function of RMPN), for larger codeword sizes, $n=5-8$, each with $M=2^{n-1}$ signaling states. We only show the low temperature transduction cases since as in the $n=3,4$ cases the JDR cannot attain a $(1-\mathcal{J})$ lower than the Helstrom limit when the transduction is performed at $T=1$K. In addition, instead of showing behavior with increasing ansatz layers, for simplicity we  show the $(1-\mathcal{J})$ achieved by the optimized $n$-qubit unitaries. For these larger codewords, we see again that there is a significant region of RMPN where the average error in decoding is less than the relevant Helstrom limit, indicating a quantum advantage.

To see how the error probability changes with increasing codeword size, in \cref{fig:codewords} we show the $(1-\mathcal{J})$ as a function of RMPN for various $n$ with $M=4$ fixed (for transduction at $1$mK). The average error probability decreases with increasing codeword size (\ie as the rate of the code decreases) and it does so appreciably in the region with the greatest quantum advantage, $\sim 0.05 <  \text{RMPN} < 0.5$.

\begin{figure} 	\includegraphics[width=0.9\linewidth]{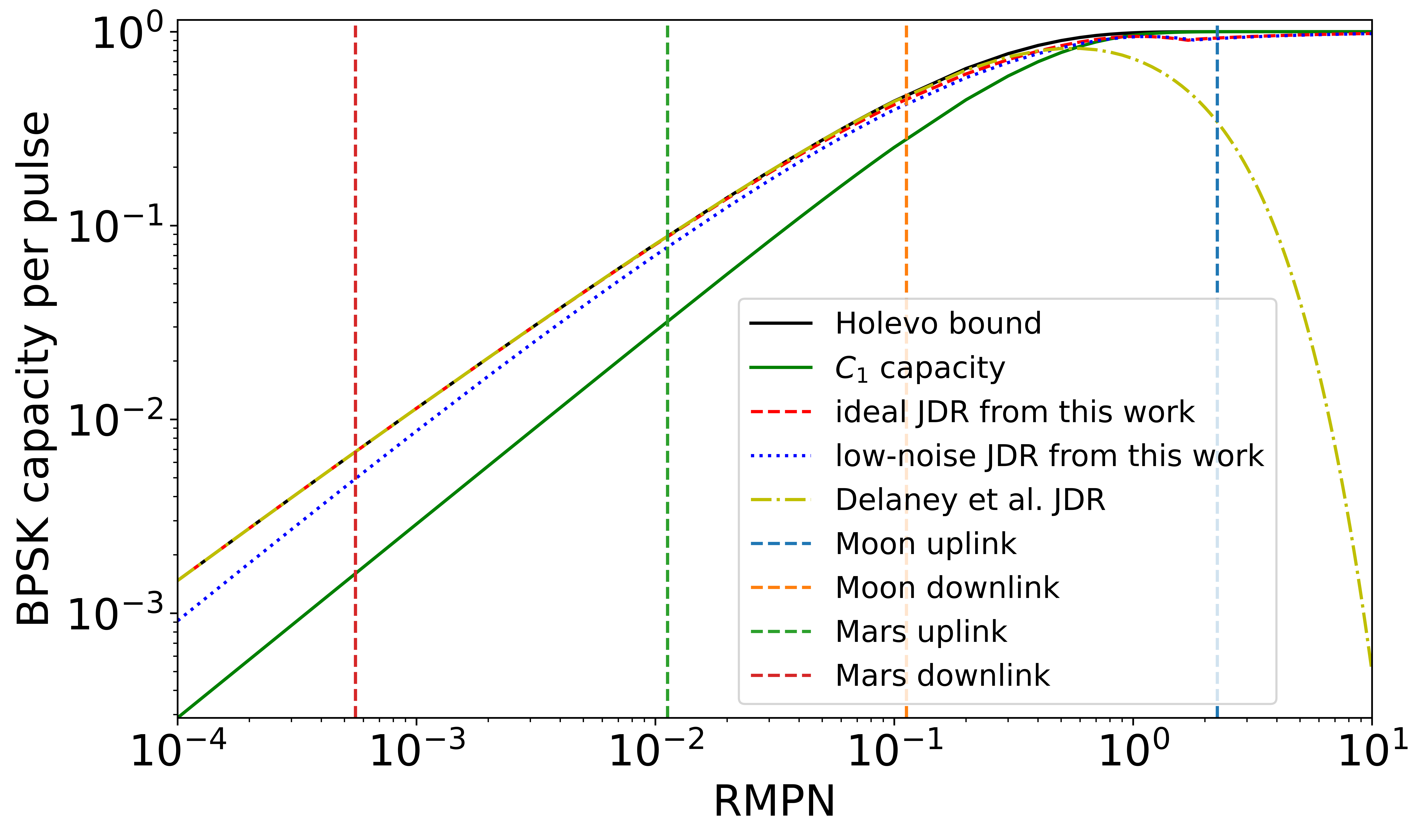}
	\caption{BPSK channel capacities per pulse as functions of RMPN. For comparison, we show our JDR performance both with ideal transduction (\(\bar{n}_{\rm tr}\) = 0, \(\eta_{\rm tr}\) = 1, red dashed curve) and in the low noise regime (T=1mK, blue dotted curve), as well as the capacity of the receiver proposed by Delaney {\etal} \cite{Delaney_2021}.  Note that both the ideal and the low-noise JDRs outperform the $C_1$ capacity until the RMPN reaches \(\sim\)0.8, a slightly larger range than for the Delaney {\etal} receiver. \label{fig:capacities}}
\end{figure}

Finally, to understand the asymptotic advantage provided by our JDR consisting of optomechanical transduction and variational quantum computation, we compare various BPSK capacities (per pulse) in \cref{fig:capacities}. We plot the capacity for our JDR in the ideal transduction case and in the low temperature transduction case. Both provide an improvement over the individual detection receiver capacity ($C_1$), by almost an order of magnitude in the very low RMPN regime. For comparison we also show the BPSK capacity of the JDR proposed in Delaney {\etal} \cite{Delaney_2021} that proceeds by probabilistic transduction into trapped ions. Our JDR in the ideal limit achieves the same capacity as that of Delaney {\etal} at most RMPNs, but is slightly below that capacity when heating and loss of low temperature transduction is factored in.  This is not surprising since the transduction model in Delaney {\etal} does not incorporate non-idealities such as thermal noise and loss. Notably, our JDR overcomes the dip in capacity at large RMPNs of the receiver in Ref. \cite{Delaney_2021}, which is caused by heralding a successful transduction on measuring zero photons, and remains close to the Holevo bound on capacity at all RMPNs.  The vertical lines in \cref{fig:capacities} show the typical RMPN for various space communication links, as calculated in Ref. \cite{Delaney_2021}.

\section{Experimental demonstration of variational circuit decoding}
\label{sec:expt}
In this section we aim to demonstrate the robustness of the quantum computer enabled receiver concept to experimental noise. The transduction model accounts for some of the noise in the transduction physics, including loss and thermal effects, and in this section we account for the noise in the quantum computation. There are some fundamental reasons to expect some robustness to noise. The first is that this application of quantum computers does not rely on scale of computation (in terms of number of qubits or circuit depth) to achieve a quantum advantage -- instead, the quantum computer is enabling a measurement that is not possible in the classical regime. Second, the aim of the computation is not an exact answer but rather a reduction in average probability of error and thus could be more tolerant to error in the circuit implementation. 

To assess the robustness we implement the trained variational circuit for codeword size $n=3$ and transduction at $T=1$mK from \cref{sec:demo} on the cloud-based IBM device \texttt{ibm\_algiers}. The initial states to the circuit are prepared using a layer of single qubit gates at the beginning of the circuit. Since the initial states -- the codeword states after transduction at $1$mK -- are mixed, for this demonstration we decompose these mixed states into pure state ensembles that are then operated on by the circuit, and then the circuit results are combined to compute the average probability of error; \ie $\bra{b_i}\mathcal{E}(\rho)\ket{b_i} = \sum_i \lambda_i \bra{b_i}\mathcal{E}(\ket{\psi_i}\bra{\psi_i}) \ket{b_i}$, where $\lambda_i$ and $\ket{\psi_i}$ are the eigenvalues and eigenvectors of $\rho$. The average error probability was estimated using $8192$ executions of the circuit (shots) per input eigenstate.

\begin{figure} 	\includegraphics[width=0.9\linewidth]{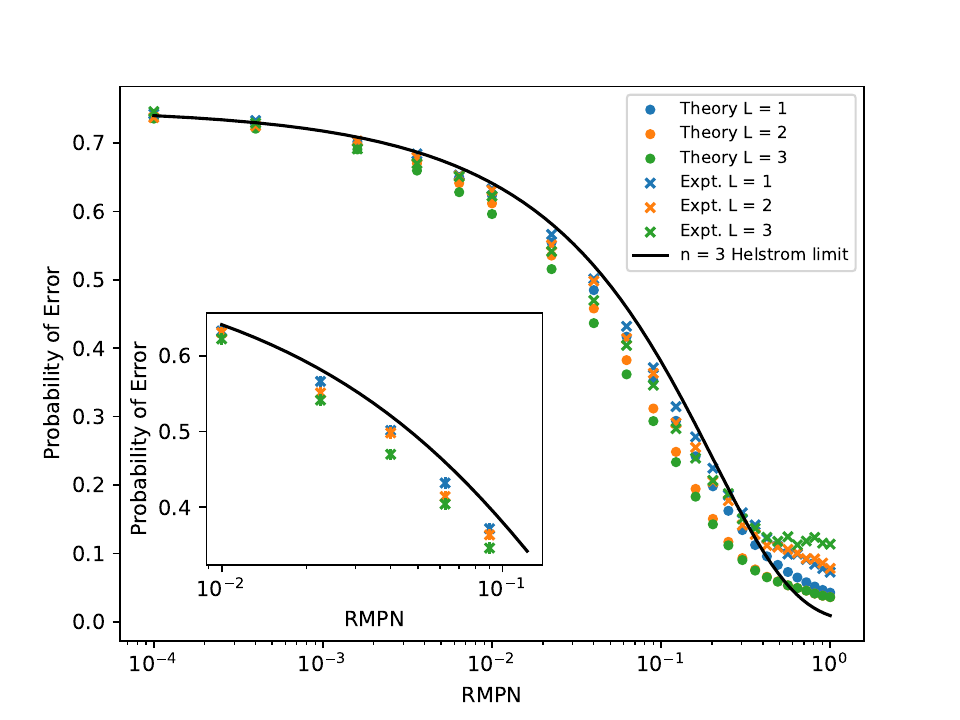}
	\caption{Average probability of error ($1-\mathcal{J}$) for decoding $M=4$ $3$-qubit codewords with an optimized variational circuit implemented on  \texttt{ibm\_algiers} ($\times$ markers). $L$ is the number of layers of the variational ansatz in \cref{fig:variational_circuit} that are optimized.  The codeword input states were initialized using the model of optomechanical transduction at $T=1$mK developed above. The black curve is the relevant Helstrom limit, and the dots are the theoretical, error-free error probabilities for the corresponding circuits. The inset displays a zoomed in view of the region of the plot with the largest quantum advantage and shows the data points with error bars. \label{fig:ibm}}
\end{figure}

\cref{fig:ibm} shows the probability of error achieved by the experimental circuit implementation as a function of RMPN for variational circuits with $L=1-3$, with the theoretical calculations and the classical limit for comparison (the latter quantities are the same as in \cref{fig:n3n4}(a)). While the experimental $(1-\mathcal{J})$ values are consistently greater than the theoretical ones, remarkably the probability of error is lower than the classical $n=3$ Helstrom limit for a large range of RMPN values. In \cref{tab:calibration} we show the device parameters reported by IBM at the time of the experiments. The parameters reveal a conventional contemporary superconducting qubit processor, with average fidelity and coherence characteristics. Thus, even with current quantum computer gate and qubit qualities, the advantage presented by a JDR can be realized if the transduction fidelity is reasonably high.

\begin{table}
    \centering
    \begin{tabular}{|c|c|c|c|}
    \hline
         Qubit&  T1&  T2&  Avg. Readout assignment error\\
         \hline
         0&  155 $\mu$s&  80 $\mu$s&  0.85\%\\
         1&  152 $\mu$s&  246 $\mu$s&  0.63\%\\
         2&  183 $\mu$s&  285 $\mu$s&  0.86\%\\
    \hline 
    \end{tabular}
    \begin{tabular}{|c|c|c|}
    \hline
    Qubit pair & CNOT error rate & CNOT gate time \\
    \hline
    $0, 1$ & 0.59\% & 260 ns\\
    $1, 2$ & 0.60\% & 320 ns\\
    \hline
    \end{tabular}
    \caption{\texttt{ibm\_algiers} device parameters for the qubits used to generate data in \cref{fig:ibm}.}
    \label{tab:calibration}
\end{table}

\cref{fig:ibm} illustrates the robustness of the JDR predictions to hardware noise. However, it should be noted that the longest circuit executed, with $L=3$ layers, contains only $6$ CNOT gates. This is sufficient to minimize average error probability in this case (as shown in \cref{fig:n3n4} the error probabilities for $L=3$ coincide with those achieved by an optimized 3-qubit unitary transformation), but for larger codeword sizes we expect to require more layers and thus more CNOT gates. For example, for $n=4$, the $L=4$ optimized circuit has $16$ CNOT gates. While this is not too large, due to the limited connectivity of the \texttt{ibm\_algiers} device the compiled circuit has $52$ CNOTs. As a result, we were unable to observe a significant quantum advantage at this codeword size using this device. It is possible that more sophisticated variational optimization techniques that take into account device connectivity and hardware noise, \eg  the approach in Ref. \cite{Cincio_2021}, will improve the experimental performance for larger codeword sizes. Finally, we have also simulated the impact of a simple depolarizing model of gate noise on the receiver circuit performance  and the results are presented in \cref{app:nm}.

\section{Conclusions and Discussion}
\label{sec:conc}
We have shown how a JDR for optical communication can be constructed from optomechanical transduction and superconducting quantum information processing devices. The performance of such a JDR depends on the transduction physics, specifically on the thermal noise introduced by the mechanical oscillator and losses in the transduction chain. We predict that operating the optomechanical transducer around $T=1$mK, and with complete tunability of the couplings between the mechanical oscillator and optical and microwave cavities, one can achieve the transduction fidelities required to demonstrate a JDR with an advantage over all classical receivers that process the received pulses one at a time. This advantage can be realized with quantum computers as small as 3 qubits and with circuits containing as few as 6 CNOT gates. Fundamentally, the advantage arises from the ability to engineer general measurements (positive operator valued measures or POVMs) on the codewords. 

In addition to numerical simulations, we implemented the variational circuit-based decoder on an IBM cloud-based quantum computer to demonstrate that even with current levels of hardware noise, if the transduction is high fidelity enough, a quantum computer-enabled JDR can surpass classical bounds on decoding error. The impact of noise can be further minimized by taking noise models into account while performing the variational optimization. 

There are several challenges to an end-to-end realization of quantum computer-enabled JDRs. The first is transduction fidelity -- as discussed in \cref{sec:trans}, state-of-the-art optomechanical transduction efficiencies fall far short of the $\eta_{\rm tr}\sim 0.9$ assumed in our calculations. Given current devices, it is not possible to surpass the performance of state-of-the-art optical receivers, that can achieve detection efficiencies of $\sim 90\%$ \cite{PhysRevLett.106.250503}. The linewidths of the optical and microwave cavities used in transducers need to be reduced, and technical sources of noise must be minimized to achieve transduction fidelities required for achieving a quantum advantage. 
A second major challenge is the integration of high quality quantum transduction with quantum computers. Despite the progress in quantum transduction in recent years \cite{Lauk_2020, Andrews_2014, Higginbotham_2018, Forsch_2020, Hease_2020, Witmer_2020, Mirhosseini_2020, Jiang_2020,Brubaker_2022, Sahu_2022} such integration has not been demonstrated to our knowledge. However, it should be noted that unlike quantum transduction for quantum networking purposes, for the JDR application we require transduction of weak coherent states and not single photons. Furthermore, the transduction can be unidirectional, optical to microwave frequencies. Both of these aspects have the potential to make transduction for JDRs easier to implement. Finally, while we focused on optomechanical transduction in this work, it would be fruitful to study other mechanisms for transduction of optical coherent states to qubits to base quantum computer-enabled receivers on.  

\acknowledgements
This material is based upon work supported by the U.S. Department of Energy, Office of Science, Office of Advanced Scientific Computing Research, under the EXPRESS program.
Sandia National Laboratories is a multimission laboratory managed and operated by National Technology \& Engineering Solutions of Sandia, LLC, a wholly owned subsidiary of Honeywell International Inc., for the U.S. Department of Energy's National Nuclear Security Administration under contract DE-NA0003525.
This paper describes objective technical results and analysis.
Any subjective views or opinions that might be expressed in the paper do not necessarily represent the views of the U.S. Department of Energy or the United States Government.

\bibliography{biblio}

\appendix

\section{Model parameters}
\label{app:parameters}
In most of this paper, simulations of the transduction model use the following values for the physical parameters:
\begin{align}
	\omega_1 &= 2\pi \times 31 {\rm THz} \nn \\
	\omega_2 &= 2\pi \times 10 {\rm Mhz} \nn \\
	\omega_3 &= 2\pi \times 10 {\rm GHz} \nn \\
	\kappa_j &= 2\pi \times 50 {\rm kHz, ~for~~} j=1,3 \nn \\
	\gamma & = 2\pi \times 500 {\rm Hz} \nn \\
	g^{\rm max}_1 &= 2\pi \times 10 {\rm Hz} \nn \\
	g^{\rm max}_3 &= 2\pi \times 5 {\rm Hz}
\end{align}
The optical frequency, \(\omega_1\), corresponds to the wavelength \(\lambda=1550\)nm, which is in the commonly used C-band of optical communication. All of these parameter values are consistent with the optomechanical transduction devices in Refs. \cite{Andrews_2014,Higginbotham_2018}, except for the cavity linewidths $\kappa_1, \kappa_3$. We choose smaller values of the latter so as to ensure operation within the strongly coupled regime, $\kappa_j \ll G_j^{\rm max}$, which means that the state is transferred from the optical to the microwave cavity at faster timescales than the cavity loss, thus ensuring high efficiency transduction \cite{Wang_Clerk_2012}. While the early optomechanical transduction experiments in Refs. \cite{Andrews_2014,Higginbotham_2018} did not use such narrow linewdith cavities, these are well within reach of modern optomechanical devices \cite{eerkens2015optical,planz2023membrane}.

We choose coherent drive amplitudes such that the maximum amplified coupling strengths in the linearized Hamiltonian, \(G_j^{\rm max}= g^{\rm max}_j|\mathcal{B}_j|\), are \(2\pi\times 1\) MHz for \(j=1,3\). Through the relationship \(|E_j| = \sqrt{\nicefrac{2 P_j \kappa_j}{\hbar \omega_j}}\), this corresponds to drive powers of \(P_1\sim 1\)W for the optical cavity and \(P_3\sim 1\)mW for the microwave cavity.

Finally, we consider two temperatures for the mechanical reservoir, \(T=1K\) and \(T=1\)mK, corresponding to \(\bar{n}\sim 2000\) and \(\bar{n}\sim 2\) average thermal occupation, respectively. We assume the initial state of the mechanical oscillator is in thermal equilibrium with its environment, and therefore \(\bar{n}_0=\bar{n}\).

\section{Sequential swap protocol}
\label{app:protocol}
Here we summarize the sequential swap protocol, initially developed in Ref. \cite{Tian_Wang_2010} for transferring a coherent state from the optical cavity to the microwave cavity, based on the linearized model developed in \cref{sec:trans}. In addition, we choose the cavity detunings such that \(\Delta_j - g_j\mathcal{Q}_2 = \omega_2\) for \(j=1,3\). This results in a linear model that describes a beamsplitter interaction between the mechanical mode and the two cavity modes.

The protocol operates in the strong coupling regime, where \(G_j \gg \kappa_j, \gamma\), and we assume that the cavity drive amplitudes, \(E_j\), have been chosen such that \(G_j\ll \omega_2\) is satisfied and we can make the rotating wave approximation to the linearized Hamiltonian (\cref{eq:lin_ham}). The initial states of the three DOFs are: vacuum for the optical and microwave cavity, and a thermal state with average occupation \(\bar{n}_0\) for the mechanical oscillator.

\noindent \emph{Step 1:} \(g_1=g_3=0\). Load all oscillators with their steady states – coherent states with amplitude \(\mathcal{B}_j\) from \cref{eq:steady_states}. This can be done by driving and waiting for the system to relax, or more practically, through resonant drives of the two cavities and parameteric displacement of the mechanical oscillator \cite{Tian_Wang_2010}. During this process, the optical cavity is also driven by the received input pulse with unknown state \(\ket{\alpha}\). At the end of this step, the idealized state of the three DOF is
\begin{align}
    \ket{\mathcal{B}_1 + \beta}\bra{\mathcal{B}_1 + \beta}\otimes \varrho(\bar{n}_0, \mathcal{B}_2) \otimes \ket{\mathcal{B}_3}\bra{\mathcal{B}_3},
\end{align}
where \(\beta = \sqrt{\eta}_{\rm in}\alpha\) and \(\varrho(\bar{m}, a)\) is a thermal state with thermal occupation \(\bar{m}\) displaced to amplitude \(a\).

\noindent \emph{Step 2:} Set \(g_1=g_1^{\rm max}, g_3=0\) and wait \(\tau_1=\nicefrac{\pi}{2G^{\rm max}_1}\), at which point the beam-splitter interaction swaps the state of the shifted oscillators. At the end of this step the idealized state of the system is
\begin{align}
\ket{\mathcal{B}_1}\bra{\mathcal{B}_1}\otimes \varrho(\bar{n}_0, \mathcal{B}_2+\beta) \otimes \ket{\mathcal{B}_3}\bra{\mathcal{B}_3}.
\end{align}

\noindent \emph{Step 3:} Set \(g_1=0, g_3= g_3^{\rm max}\) and wait \(\tau_3=\nicefrac{\pi}{2G^{\rm max}_3}\), to execute a second swap into the microwave cavity. The idealized state of the system after this step is
\begin{align}
\ket{\mathcal{B}_1}\bra{\mathcal{B}_1}\otimes \varrho(\bar{n}_0, \mathcal{B}_2) \otimes \ket{\mathcal{B}_3+\beta}\bra{\mathcal{B}_3+\beta}.
\end{align}

\begin{figure*}[t]
\includegraphics[width=1\linewidth]{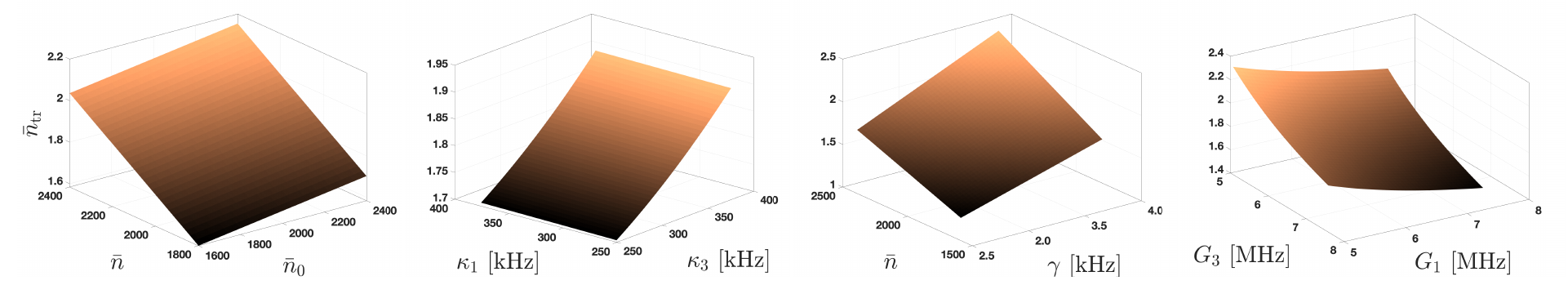}
\caption{\label{fig:n_tr} Variation of \(\bar{n}_{\rm tr}\) as physical parameters in the transduction model are varied from their fiducial values given in \cref{app:parameters}. All parameters except the two varied in each plot are held at their fiducial values. The temperature is assumed to be \(T=1\)K, except for the cases where \(\bar{n}\) is varied. }
\end{figure*}

We have shown the idealized states after each step in the transfer protocol above, but these do not take into account the dissipative and heating processes on the three DOF during the transfer time \(\tau_1+\tau_3\). Taking these into account results in a final state in the microwave cavity that is a displaced thermal state, \cref{eq:disp_thermal} in the main text. The loss and heating parameters in that state are derived in Ref. \cite{Wang_Clerk_2012}, and in our notation are:
\begin{align}
	\eta_{\rm tr} &= e^{-\vartheta_1\tau_1 - \vartheta_3\tau_3}, \nn \\
	\bar{n}_{\rm tr} &= \frac{1}{2}\Big(e^{-(\vartheta_1\tau_1+\vartheta_3\tau_3)}(1+\nu_1^2) + \kappa_3\beta_3 + \gamma(2\bar{n}+1)\mu_3 \nn \\
	&~~~~~ +e^{-\vartheta_3\tau_3}(\nu_3^2 +\kappa_1\mu_1+\gamma(2\bar{n}+1)\alpha_1)-1\Big),
\end{align}
where
\begin{align}
	\vartheta_j &= \nicefrac{(\kappa_j+\gamma)}{2}, \nn \\
	\nu_1 &= \nicefrac{(\kappa_1-\gamma)\sqrt{2\bar{n}_0 +1}}{4G^{\rm max}_1}, \nn \\
	\nu_3 &= \nicefrac{(\kappa_3-\gamma)}{4G^{\rm max}_3}, \nn \\
	\alpha_1 &\approx \int_0^{\tau_1} d\tau (\cos(G^{\rm max}_1\tau) + \nu_1\sin(G^{\rm max}_1\tau))^2 e^{-\nicefrac{\vartheta_1\tau}{2}}, \nn \\
	\beta_3 &\approx \int_0^{\tau_3} d\tau (\cos(G^{\rm max}_3\tau) - \nu_3\sin(G^{\rm max}_3\tau))^2 e^{-\nicefrac{\vartheta_3\tau}{2}}, \nn \\
	\mu_j &\approx \int_0^{\tau_j} d\tau \sin^2(G_j^{\rm max}\tau)e^{-\kappa_j\tau}. \nn
\end{align}

While this loss parameter is easy to interpret, the heating parameter, \(\bar{n}_{\rm tr}\) (interpreted as the number of thermal photons added by the transduction process), has a complicated dependence on the parameters. To illustrate the behavior of \(\bar{n}_{\rm tr}\), in \cref{fig:n_tr} we plot it as a function of some of the physical parameters as they are varied from their fiducial values at \(T=1\)K. The overall variation is similar for the lower temperature of \(T=1\)mK. As seen from these plots, the variation of \(\bar{n}_{\rm tr}\) with the physical parameters is mild.

\cref{tab:params} in the main text shows the values of \(\bar{n}_{\rm tr}\) and \(\eta_{\rm tr}\) at the fiducial transduction parameter values given in \cref{app:parameters} and at the two possible operating temperatures.

\section{Tuning the Jaynes-Cummings interaction}
\label{sec:JC}

In this appendix we present details of the Jaynes-Cummings (JC) interaction dynamics that form the basis of the transduction from microwave cavity state to qubit state.

We can understand the microwave cavity to qubit transduction step by using the exact solution to the JC interaction in the Heisenberg picture \cite[Ch. 6.2.2]{scully_zubairy_1997}, which for the atomic lowering operator takes the form:
\begin{align}
	\sigma^-(t) = e^{iCt}\Big{[}\Big{(}\cos{\kappa t} + iC\frac{\sin{\kappa t}}{\kappa}\Big{)}\sigma^-(0) - {i\chi} \frac{\sin{\kappa t}}{\kappa} b_3(0)\Big{]},
\end{align}
where \(\kappa = \chi\sqrt{N+1}\), \(N = b_3\dg b_3 + \sigma^+\sigma^-\), and \(C = \chi(b_3\sigma^+ + b_3\dg\sigma^-)\) are constants of motion. The two quantities $\expect{\sigma^+(t)\sigma^-(t)}$ and $\expect{\sigma^-(t)}$ completely determine the qubit state at any time. Considering BPSK signaling, the two qubit reduced density operators that correspond to signaling states $\ket{\pm \beta}$ are
\begin{align}
	\rho^{\pm}_q(t) = \begin{bmatrix}
  \expect{\sigma^+(t)\sigma^-(t)}_{\pm} & \expect{\sigma^-(t)}_\pm\\
  \expect{\sigma^+(t)}_\pm & \expect{\sigma^-(t)\sigma^+(t)}_\pm
  \end{bmatrix},
\end{align}
where, \emph{e.g.,} \(\expect{\sigma^-(t)}_\pm = \tr{([\rho^{\pm}_f(0)\otimes\rho_q(0)]\sigma^-(t))}\), with $\rho_q(0) = \ket{g}\bra{g}$ and $\rho^{\pm}_f(0) = \varrho(\bar{n}_{\rm tr},\pm \sqrt{\eta_{\rm tr}} \beta)$.

Using $b_3(0)\ket{\alpha} = \alpha \ket{\alpha}$ and $\sigma^-(0)\ket{g} = 0$, the off-diagonal expectation value can be evaluated to the simplified expression
\begin{align}
\expect{\sigma^-(t)}_{\pm} = -\frac{i\chi}{\pi \bar{n}_{\rm tr}}\int {\rm d}^2\alpha~ & e^{-\frac{|\alpha \mp \sqrt{\eta_{\rm tr}}\beta|^2}{\bar{n}_{\rm tr}}} \alpha \nn \\
&\bra{g \alpha} \cos Dt \frac{\sin \kappa t}{\kappa} \ket{g \alpha},  \nn
\end{align}
where $D \equiv \chi \sqrt{N}$ and $\ket{g\alpha}$ is shorthand for $\ket{g}\otimes \ket{\alpha}$. Expanding the operators $\kappa$ and $D$ in the Fock basis and simplifying:
\begin{align}
	\expect{\sigma^-(t)}_{\pm} =& \mp i\frac{\sqrt{\eta_{\rm tr}}\beta e^{-\frac{\eta_{\rm tr}|\beta|^2}{\bar{n}_{\rm tr}+1}}}{(\bar{n}_{\rm tr}+1)^2} \nn \\
	& \sum_{n=0}^\infty \frac{\cos(\chi t \sqrt{n}) \sin(\chi t \sqrt{n+1})}{\sqrt{n+1}} f_n(\eta_{\rm \tr} |\beta|^2, \bar{n}_{\rm tr}),
	\label{eq:sigminus}
\end{align}
where the function $f_n$ is a summation involving its arguments whose exact form is not important for our analysis below, except for the fact that it satisfies $f_n\geq 0, \forall n$. Note that the only dependence on $\arg(\beta)$ in this expression is from the $\beta$ prefactor, and hence we can say that the qubit phase in the $X-Y$ plane is a simple function of the phase of the coherent state, \ie $\arg \expect{\sigma^+(t)} = i \arg (\beta^*)$.

The distinguishability of the transduced qubit states is evaluated by computing their trace distance. By simplifying the expression for $\expect{\sigma^+(t)\sigma^-(t)}_\pm$ using the above observations we can see that this quantity does not depend on $\arg (\beta)$, and therefore the trace distance between the two qubit states transduced from BPSK signals is
\begin{align}
\tau \equiv \frac{1}{2}\tr |\rho^+_q(t) - \rho_q^-(t)| = |\expect{\sigma^-(t)}_+ - \expect{\sigma^-(t)}_-|.
\end{align}
Using \cref{eq:sigminus}, this can be expanded to
\begin{align}
\tau = 2\frac{\sqrt{\eta_{\rm tr}}|\beta| e^{-\frac{\eta_{\rm tr}|\beta|^2}{\bar{n}_{\rm tr}+1}}}{(\bar{n}_{\rm tr}+1)^2}
	 \Big\vert \sum_{n=0}^\infty \frac{\cos(\chi t \sqrt{n}) \sin(\chi t \sqrt{n+1})}{\sqrt{n+1}} f_n \Big\vert.
	 \label{eq:tau}
\end{align}
Examining this quantity, and approximating the arguments to the two trigonometric functions as the same, it is clear that a strategy for maximizing it is to choose $t = \tilde{t} \sim \frac{\pi}{4\chi \sqrt{N}}$, where $N$ is roughly the number of photons in the state $\varrho(\bar{n}_{\rm tr}, \pm \sqrt{\eta}_{\rm tr}\beta)$. Despite this guide, unfortunately, the optimal time does not have a simple expression. However, we can numerically evaluate it and when this is done we find it has a complex dependence on the parameters and a dependence that is a function of the mean photon number in the microwave cavity, $MPN \equiv |\sqrt{\eta_{\rm tr}}\beta|^2+\bar{n}_{\rm tr}$.
If we focus on the short-time regime, $0 \leq t \leq 5/\chi$, then the optimal time takes the form
\begin{align}
t^* \approx \frac{\pi}{4\chi\sqrt{|\sqrt{\eta_{\rm tr}}\beta|^2+1.5\bar{n}_{\rm tr}}},
\label{eq:topt_fit}
\end{align}
for values of $MPN\geq \nicefrac{1}{4}$. This form follows that of the optimal time identified by the analytical arguments above, $\tilde{t}$. However, for values of $ 0 < MPN \leq \nicefrac{1}{6}$, we find that the optimal time is simply
\begin{align}
t^* \approx \frac{\pi}{2\chi}.
\label{eq:topt_pi2}
\end{align}
In this regime, the fact that the two trigonometric functions in \cref{eq:tau} have different $n$ dependence ($\sqrt{n}$ versus $\sqrt{n+1}$) means that the above argument for maximizing $\tau$ is not valid -- in this regime the $n=0$ term dominates the sum and we should simply maximize $\sin(\chi t)$. In the intermediate regime, $\nicefrac{1}{6} < MPN < \nicefrac{1}{4}$, the optimal time crosses over from \cref{eq:topt_pi2} to \cref{eq:topt_fit}, and this crossover depends on the balance between the coherent photons ($|\sqrt{\eta_{\rm tr}}\beta|^2$) and thermal photons ($\bar{n}_{\rm tr}$) in the state.

Finally, we mention that if we go beyond the short-time regime, it is possible in some regimes of $MPN$ to obtain maximum trace distance at \(t\approx 8/\chi\) \cite{Han_2018}. We generally do not consider such long time dynamics in this work though, since (i) the gain in trace distance at long times is only slight, and (ii) the short-time regime is most relevant to our application since we want to transduce the information to the qubits as quickly as possible to increase bandwidth and also to minimize the impact of decoherence processes. In almost all of the numerical calculations of the JC interaction and qubit state used in the main text, we optimize the interaction time in the range $0 \leq t \leq 5/\chi$. The exception is in the capacity plot of \cref{fig:capacities}, where we optimize over a longer time window, $0 \leq  t \leq 10/\chi$, to capture the true optimum qubit states. 

We conclude this section with a comment on the initial state for the transduction dynamics. We have assumed that the initial state of the qubit is $\ket{g}$ throughout. This is a natural initial state to consider because it is the ground state of the qubit. Moreover, there are reasons to believe this initial state is optimal for transducing phase information from a mode using the JC interaction. For example, if the qubit is initialized in a superposition, the populations will depend on the phase relation between the optical phase and initial qubit phase, and in general the transduced BPSK states have no symmetry on the Bloch sphere. Therefore the distinguishability of the transduced states is not solely determined by properties of the microwave mode. We have also numerically verified that the distinguishability is not increased by choosing a different initial qubit state.

\section{Receiver performance under simulated circuit error model}
\label{app:nm}
\label{sec:NM}
\begin{figure}[t]
\includegraphics[width=1\linewidth]{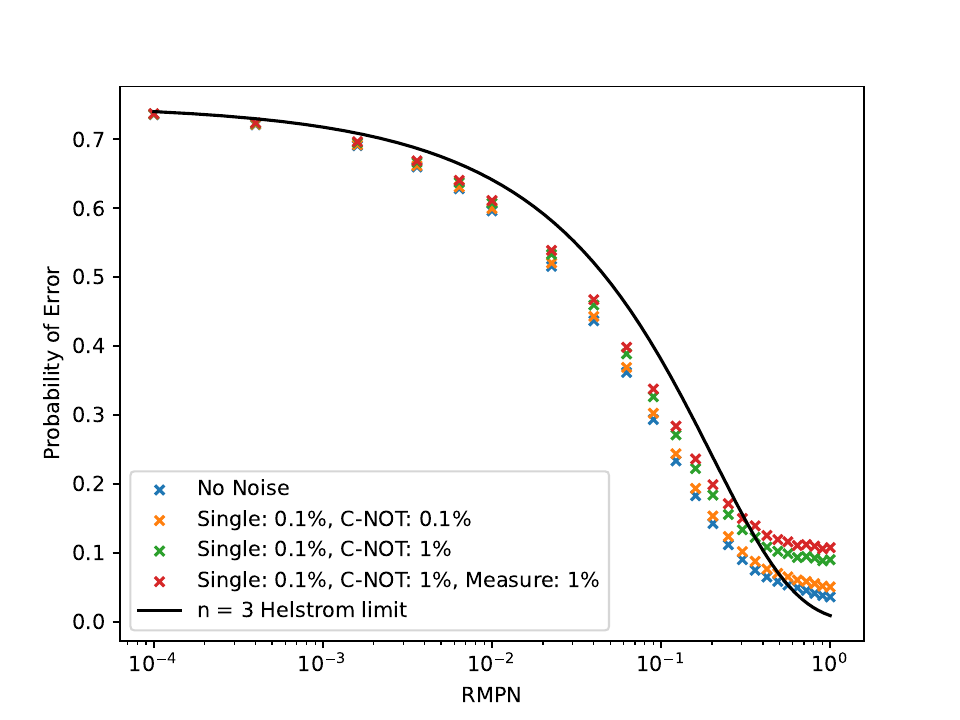}
    \caption{Probability of error for decoding 4 3-qubit codewords transduced at $T=1$mK with the 3 layer variational circuit under the theoretical error model described in the text. Blue is the case with no noise. Orange is the case with $p_1=0.1\%, p_2=0.1\%, p_m=0$. Green is the case with $p_1=0.1\%, p_2=1\%, p_m=0$. Red is the case with $p_1=0.1\%, p_2=1\%, p_m=1\%$.
    \label{fig:noise}}
\end{figure}

In this appendix, we study the performance of the joint receiver's variational quantum circuit under a simple theoretical noise model. In \cref{fig:noise}, we plot the probability of error in decoding under varying levels of noise for the case of 4 3-qubit codewords transduced at  $T=1$mK with the 3 layer variational circuit used in the main text. The noise model consists of (1) a single-qubit depolarizing channel after every single qubit gate with error probability $p_1$, (2) a two-qubit depolarizing channel after every CNOT gate with error probability $p_2$, and (3) a measurement error probability of $p_m$.

As expected, the performance degrades as a function of increasing noise, but the quantum advantage persists for intermediate RMPNs even at appreciable levels of depolarizing noise (e.g., 0.1\% error on single qubit gates, 1\% error on CNOT gates and measurements). This is largely a consequence of the short depth variational circuit required to perform codeword discrimination.

\end{document}